\documentclass[aps,prl,reprint,groupedaddress,longbibliography]{revtex4-2}

\usepackage[utf8]{inputenc}
\usepackage{amsmath,amssymb}
\usepackage{amsfonts}
\usepackage{slashed}
\usepackage{xcolor}

\usepackage[pdftex,hypertexnames=false]{hyperref}
\usepackage{bbm}
\hypersetup{ 
pdfstartview=FitV,
colorlinks=true, 
linkcolor=black, 
citecolor=blue, 
filecolor=black, 
urlcolor=magenta,
}
\usepackage{graphicx}
\usepackage[caption=false]{subfig}
\usepackage{ulem}

\begin{document}
	\preprint{CPHT-RR011.022022}
	\title{Destroying superconductivity in thin films with an electric field}
	\author{Andrea Amoretti$^{\, 1,2}$, Daniel K. Brattan$^{\, 1,3}$, Nicodemo Magnoli$^{\, 1,2}$, Luca Martinoia$^{\, 1,2}$, Ioannis Matthaiakakis$^{\, 1,2}$, and Paolo Solinas$^{\, 1,2}$.}
	\affiliation{\textbf{1} Dipartimento di Fisica, Universit\`a di Genova,
		via Dodecaneso 33, I-16146, Genova, Italy,
	}
	\affiliation{\textbf{2} I.N.F.N. - Sezione di Genova, via Dodecaneso 33, I-16146, Genova, Italy,}
	\affiliation{\textbf{3} CPHT, \'{E}cole Polytechnique, 
		91128 Palaiseau cedex, France.}
	
	\email{andrea.amoretti@ge.infn.it}
	\email{danny.brattan@gmail.com}
	\email{nicodemo.magnoli@ge.infn.it}
	\email{luca.martinoia@ge.infn.it}
	\email{ioannis.matthaiakakis@edu.unige.it}
	\email{paolo.solinas@ge.infn.it}
	\begin{abstract}
		In this paper we use a Ginzburg-Landau approach to show that a suitably strong electric field can drive a phase transition from a superconductor to a normal metal. The transition is induced by taking into account corrections to the permittivity due to the superconductive gap and persists even when screening effects are considered.  We test the model against recent experimental observations in which a strong electric field has been observed to control the supercurrent in superconducting thin films. We find excellent agreement with the experimental data and are able to explain several observed features. We additionally suggest a way to test our theoretical proposal via superconductor-superconductor electron tunneling. 
	\end{abstract}
	
	\maketitle
	
	While the effects of a magnetic field in superconductors have been thoroughly studied, little has been done
	regarding static electric fields. The belief is that near-complete screening effects in metallic superconductors make the electric field irrelevant to superconductivity \cite{ashcroft1976solid}. Recently however, the authors of \cite{DeSimoni_2018} reported field-effect control of the supercurrent in thin, all-metallic transistors made of different Bardeen-Cooper-Schrieffer (BCS) superconductor
	\cite{tinkham2004introduction,kopnin2001theory} films. They found that, at low temperature, these transistors
	presented a monotonic decay of the critical current under increasing electrostatic field. 
	For strong enough electric fields normal metal behaviour was observed.
	This phenomenon is known as the Superconductive Field Effect (SFE) and the results have been recently confirmed by other experimental groups \cite{Ritter_2021,Alegria_2021,Gokolenkov_2021,Orus_2021,Basset2021,Ritter2021_arxiv, Elalaily2021}.
	
	The microscopic origin of this phenomenon is unclear. All the materials \cite{DeSimoni_2018,Ritter_2021,Alegria_2021,Gokolenkov_2021,Orus_2021,Basset2021,Ritter2021_arxiv, Elalaily2021}  analyzed are well described by the conventional BCS theory and metallic in the normal phase (hence it is surprising that the electrostatic field could play any role). It has been suggested that energetic quasi-particle injection from the gate control \cite{Alegria_2021,Gokolenkov_2021} or energy or phase fluctuations \cite{Basset2021} could be responsible for the observations. These ideas seem to be precluded by recent ionic-gating experiments, where the electric field is generated by crystallised charges \cite{Paolucci2021_ionic}, as there is no moving charge.
	
	Alternate microscopic proposals have been suggested to explain the experimental results of \cite{DeSimoni_2018,Ritter_2021,Alegria_2021,Gokolenkov_2021,Orus_2021,Basset2021,Ritter2021_arxiv, Elalaily2021}. These include electric-field induced spin-orbit polarization \cite{Mercaldo_2020}, Rashba-like surface effects \cite{Chirolli_2021} and the excitation of an exotic superconducting state due to Schwinger-like effects \cite{Solinas_2021}. All these proposals lead to a weakening of superconductivity but they do not fully explain other experimental results \cite{DeSimoni_2018,Paolucci_2018,Paolucci_2019,Paolucci_2019_simm,Paolucci_2019_squid,Paolucci2021_ionic} . 
	
	Given that the SFE persists in many different compounds with different experimental setups and geometries \cite{Paolucci_2018,Paolucci_2019,Rocci_2020,Bours_2020, Puglia_2020,DeSimoni_2020, Rocci_2020_1} we suggest instead that the phenomenon is a property of including the electric field in standard BCS theories. Moreover, we argue that there is a phase transition at strong electric fields to a normal metal.
	
	\begin{figure}
		\centering	
		\includegraphics[width=1\linewidth]{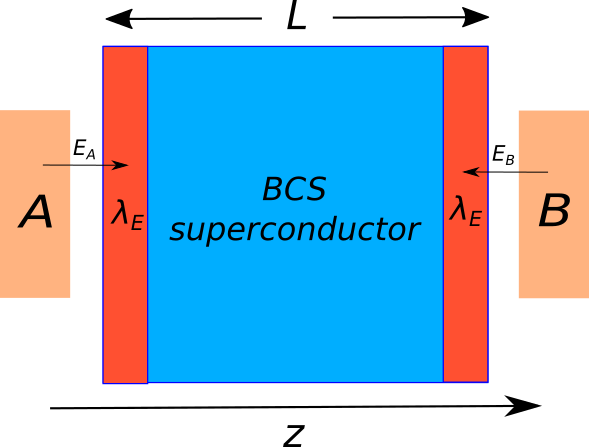}
		\caption{Section of the setup. The superconductive film is extended in the $x$ and $y$ direction and has thickness $L$ along the $z$ direction. The electric field $E_{z}$ pointing along the direction of the small black arrows is applied to both sides of the film, by means of two charge distributions $A$ and $B$, taken to be equal. The penetration length is $\lambda_E \simeq 1$ nm.}
		\label{Fig:setup}
	\end{figure}
	
	To overcome the difficulties with these microscopic approaches while still comparing our proposal with experiments, we will consider the simplest phenomenological description of a superconductor, namely, the Ginzburg-Landau (GL) mean field theory. The strength of the GL approach is twofold; we can calculate the main experimental observables and reproduce all features measured in experiment \cite{DeSimoni_2018,Alegria_2021,Ritter_2021}, and we can constrain potential microscopic explanations. We investigate an extension of the GL free energy in which the electric permittivity is modified in order to include couplings between the electric field and the order parameter. We demonstrate that, despite screening effects, a suitably strong electric field leads to a superconductor-normal metal phase transition for thin films of order the coherence length.
	
	\emph{The model \textemdash} We will analyse the setup depicted in Figure \ref{Fig:setup}, a superconductive thin film extended in the $x$ and $y$ directions and with thickness $L$ in the $z$ direction. The electric field is applied along the $z$ direction by means of two charge distributions ($A$ and $B$ in Figure \ref{Fig:setup}) placed in the vicinity of the surfaces of the film. This setup is similar to the one realized in the experiments \cite{Paolucci_2019_simm}, where the interested reader can find further technical details that do not alter the present analysis. In what follows the charges $A$ and $B$ are assumed to be equal. This implies that the direction of the electric field is reversed as we cross the middle of the channel at $z = L/2$ (see Fig.~\ref{Fig:setup}). The physical results do not depend appreciably on the sign of the charges and on such a symmetric configuration \cite{Paolucci_2019_simm}  whose purpose is simplifying numerical computations. We include screening effects in our model by assuming that they are set at the level of the normal metal phase.  Due to the reduced thickness of the sample we assume the electric field generated by the charges $A$ and $B$ inside the material to be along the $z$-direction. Hence we fix the applied electric field $\vec{E}$ to be non-dynamical with a decreasing exponential profile at the ends ($z=0$ and $z=L$) of the film i.e.
	\begin{multline} \label{Eq:electric}
		\vec{E} = - \partial_{z} \varphi(z) \hat z = E_A(z)\hat z+E_B(z) \hat z=\\
		E_0 e^{-\frac{z}{\lambda_E}} \hat z-E_0 e^{-\frac{L-z}{\lambda_E}}\hat z=	2 E_0 e^{-\frac{L}{2 \lambda_{E}}} \sinh \left( \frac{ \frac{L}{2}-z}{\lambda_{E}} \right) \hat z \; . 
	\end{multline}
	 We take the penetration length, $\lambda_E$, to be $\simeq 1$ nm  which is compatible with normal metals \cite{ashcroft1976solid}.  Note that the above formula for the electric field implies that the sample is charged, this however is an artifact of the simple screening model, as we are considering a charged-neutral film. Since the film is much larger than the penetration length, the spurious charge distribution in the bulk that would make it overall neutral will give rise to negligible effects and we consequently ignore it.
	
	Now consider the mean field description \cite{tinkham2004introduction,kopnin2001theory} of this configuration. We use the usual complex scalar superconductive order parameter written in polar form $\Psi(\vec{r}) = \Delta(\vec{r}) \exp{[i \theta(\vec{r})]}$ where $ \Delta(\vec{r})$ and $\theta(\vec{r})$ are the amplitude and the phase of the order parameter, respectively. Having fixed the electric field profile, we focus on time independent configurations. Consequently,  the resulting time-independent GL free energy is
	\begin{eqnarray}
		\label{Eq:Freenergy}
		F
		&=& \int d^{3}r \; \Big \{\frac{\hbar^2}{2m} \| \vec{\partial} \Delta \|^2 + \frac{\hbar^2}{2m} \Delta^2 \| \vec{\partial} \theta \|^2 + \left( \frac{\alpha_{2}}{2!} + q \varphi \right) \Delta^2 \nonumber \\
		&\;& \hphantom{\left. \int d^{3}x \; \right.} +  \frac{\alpha_{4}}{4!}  \Delta^4+ \frac{\epsilon[\Delta]}{2} \left( \frac{d \varphi}{dz} \right)^2 \Big \}
	\end{eqnarray}
	where $m,~q$ are the mass and charge of the Cooper pair respectively, $\varphi(z)$ is defined in \eqref{Eq:electric} and $\left\| \vec{x} \right\|$ is the vector norm of $\vec{x}$. 
	
	Our free energy \eqref{Eq:Freenergy} differs from the usual GL free energy \cite{tinkham2004introduction,kopnin2001theory} by allowing the electric permittivity to depend on the condensate density $\Delta$.  We will assume the following functional dependence on $\Delta$:
	\begin{eqnarray}
		\label{Eq:Permittivitydependence}
		\epsilon[\Delta] = \epsilon_{0} \left( 1 + \beta_{1} \Delta^2 + \beta_{2} \Delta^{4} + \ldots \right) \; ,
	\end{eqnarray}
	where $\beta_1$ and $\beta_2$ are phenomenological parameters. These additional couplings respect all the symmetries of the system and are at most quartic in the gap and quadratic in derivatives. It is consequently natural to include them within a GL functional approach. The conditions for the expansion (\ref{Eq:Permittivitydependence}) to be consistent are $\beta_1 \Delta_0^2 \ll 1$, $\beta_2\Delta_0^4 \ll 1$ where $\Delta_0^{2} = -3\alpha_2/2\alpha_4$ is a homogeneous condensate density in absence of an electric field. 
	
	Analogous terms in the electric permittivity \eqref{Eq:Permittivitydependence} already occur in the BCS context \cite{Prange_1963} due to perturbative loop corrections around the (constant, spatially independent) BCS ground state. The same perturbative approach has been followed in \cite{Virtanen_2019}. We however consider a new, inhomogeneous (coordinate dependent) ground state generated by solving the full GL equations in the presence of an external electric field.   Moreover, we promote the corrections to the permittivity found in \cite{Prange_1963} to be interactions of the bare GL functional \eqref{Eq:Freenergy}. This corresponds to finding a complete non-perturbative solution to the Ginzburg Landau equation including interactions between the order parameter and the electric field. Qualitatively our results agree with \cite{Prange_1963, Virtanen_2019} when the applied electric field is sufficiently small. Moreover, in what follows we will prove that even parametrically small coupling constants $\beta_1$ and $\beta_2$ have dramatic consequences on the order parameter $\Delta$, when the external electric field cannot be treated as a perturbative correction (as in \cite{Prange_1963}).
	
	\emph{Driving the phase transition \textemdash} The minimal energy solutions for the equations of motion resulting from \eqref{Eq:Freenergy} have constant phase $\theta$  as can be seen by examining the equation of motion,
	\begin{displaymath}
		\vec{\partial} \cdot \left( \Delta^2 \vec{\partial} \theta \right) = 0 \; ,
	\end{displaymath}
and minimizing the contribution of the second integrand in \eqref{Eq:Freenergy} to the free energy. The equation for $\Delta$ then reduces to
	\begin{eqnarray}\label{Eq:gapequation}
		\frac{\hbar^2}{m} \vec{\partial}^2 \Delta &-&  \left( \alpha_{2} + 2 q \varphi + \epsilon_{0} \beta_{1} \left( \frac{d \varphi}{dz} \right)^2 \right) \Delta \nonumber \\
		&-& \left( \frac{\alpha_{4}}{3!} + 2 \epsilon_{0} \beta_{2} \left( \frac{d \varphi}{dz} \right)^2 \right) \Delta^3=0 \; .  
	\end{eqnarray}
	This equation has to be solved imposing the usual boundary conditions at the edges of the material $\Delta'(0)=\Delta'(L)=0$ \cite{kopnin2001theory}. The $z$-dependent profile for the electric field in \eqref{Eq:electric} makes the solution of \eqref{Eq:gapequation} depend on $z$ as well. Since we are interested in global observables (e.g. the critical current of the full thin film) the relevant quantity will be the averaged gap $\Delta_{\mathrm{av.}}=\frac{1}{L} \int_0^L dz~ \Delta(z)$.  We refer to the Appendix for technical details about the numerical solution of equation \eqref{Eq:gapequation}.
	
	To have a direct comparison with the experiments, we consider the devices used in \cite{DeSimoni_2018}.
	We take $T_{\mathrm{c}} \simeq 410$ mK, London penetration length  $\lambda_L\simeq900~$nm and the coherence length $\xi_0 \simeq 100~$nm. We also assume the usual GL temperature scalings for the relevant parameters, so that $\alpha_2= -\mathcal{K} (1-T/T_{\mathrm{c}})$  with $\mathcal{K} = 6.104 \times 10^{-25}~$kg m$^{2}$ s$^{-2}$ and $\alpha_4= 6.356 \times 10^{-50}$ kg m$^{5}$ s$^{-2}$.
	
	The parameters $\beta_1$ and $\beta_2$ in \eqref{Eq:Permittivitydependence} are phenomenological, and, in order to fix them, we rely on experimental observations \cite{DeSimoni_2018}. To qualitatively understand their effect, suppose we can ignore the kinetic energy. This is certainly true in films thin enough such that the electric field penetrates deeper in the bulk  (see plots of the kinetic energy in fig.~\ref{Fig:data3} in the Appendix  for proof of concept). In this case one can define new effective GL parameters $\tilde{\alpha_2}$ and $\tilde{\alpha_4}$ averaged over the system: $\tilde{\alpha}_2 = \alpha_2 + 2 q \varphi_{\mathrm{av.}} + \epsilon_{0} \beta_{1} E^2_{\mathrm{av.}}$ and $\tilde{\alpha}_4 = \alpha_4+ 12 \epsilon_{0} \beta_{2}E^2_{\mathrm{av.}}$, where $\varphi_{\mathrm{av.}}$ and $E_{\mathrm{av.}}$ are the space averages of the scalar potential and the electric field respectively. As in the standard GL model, the phase is consequently determined by the sign of $\tilde{\alpha}_2$ (negative for superconducting, positive for metal) \cite{tinkham2004introduction}. If $\beta_{1}>0$ is chosen carefully, since $\alpha_{2}<0$, we can set $\tilde{\alpha}_{2}=0$ for some critical averaged electric field ${E}^{\mathrm{c}}_{\mathrm{av.}}$, which determines the critical point.
	Similarly, $\tilde{\alpha}_4$ affects the form of the potential energy (e.g. position of minima) and, therefore, the dependence of $\Delta$ on $E_{z}$. It does not, however, affect the value of ${E}^{\mathrm{c}}_{\mathrm{av.}}$.
	
	Focusing on samples with thickness of the order of the coherence length ($L\simeq \xi_0$), from the experimental data \cite{DeSimoni_2018}, it turns out that $\beta_1$ has a linear dependence on $T$ (see fig.~\ref{Fig:data1}b Appendix),
	\begin{equation}\label{Eq:beta1}
		\beta_1= \left[ A+ B  \left(1-\frac{T}{T_\mathrm{c}} \right) \right] \mathrm{\; m^3} \ ,
	\end{equation}
	with $A = 1.208 \times 10^{-30}$ and $B=  5.947\times 10^{-28}$. The presence of the positive constant $A$ ensures that in the limit of $T \rightarrow T_{\mathrm{c}}$, $\beta_1$ remains positive. This means that when the superconductor is in the normal phase but close to $T_{\mathrm{c}}$, the electric field cannot induce a phase transition to the superconducting state. With this information \eqref{Eq:beta1} we can estimate the correction to the critical temperature due to the $\beta_{1}$ coupling. In particular, $\tilde{\alpha}_2(0) = (1-T/T_{\mathrm{c}}) (- \mathcal{K} +  \epsilon_{0} B ({E}^{\mathrm{c} }_{\mathrm{av.}})^2 ) +   \epsilon_{0} A ({E}^{\mathrm{c}}_{\mathrm{av.}})^2$.
	Assuming ${E}^{\mathrm{c}}_{\mathrm{av.}} \sim 10^8$ V/m as in \cite{DeSimoni_2018}, we can  solve the above equation for $\tilde{\alpha}_2 = 0$, obtaining a new critical temperature $T_{\mathrm{c},new}$ and a variation $\Delta T_{\mathrm{c}} = T_{\mathrm{c},new}-T_{\mathrm{c}} \propto 3-4$ mK. This is a rough estimate but is compatible with the full numerical simulations and the absence of variation of $T_{\mathrm{c}}$ observed in the experiments  \cite{Paolucci_2018}. 
	
	Finally, the second parameter $\beta_2$, is obtained as the best fit of experimental data. As it is a best fit parameter and the temperature range of the experimental data is small, it is difficult to extract a precise temperature dependence. Nevertheless we find that this parameter has a mild temperature dependence, with $-4 \times 10^{-54}  \mathrm{\; m^6} \lesssim \beta_{2} \lesssim -6 \times 10^{-54}  \mathrm{\; m^6}$ for the range of temperatures examined.
	
	\begin{figure}
		\centering	
		\includegraphics[width=1\linewidth]{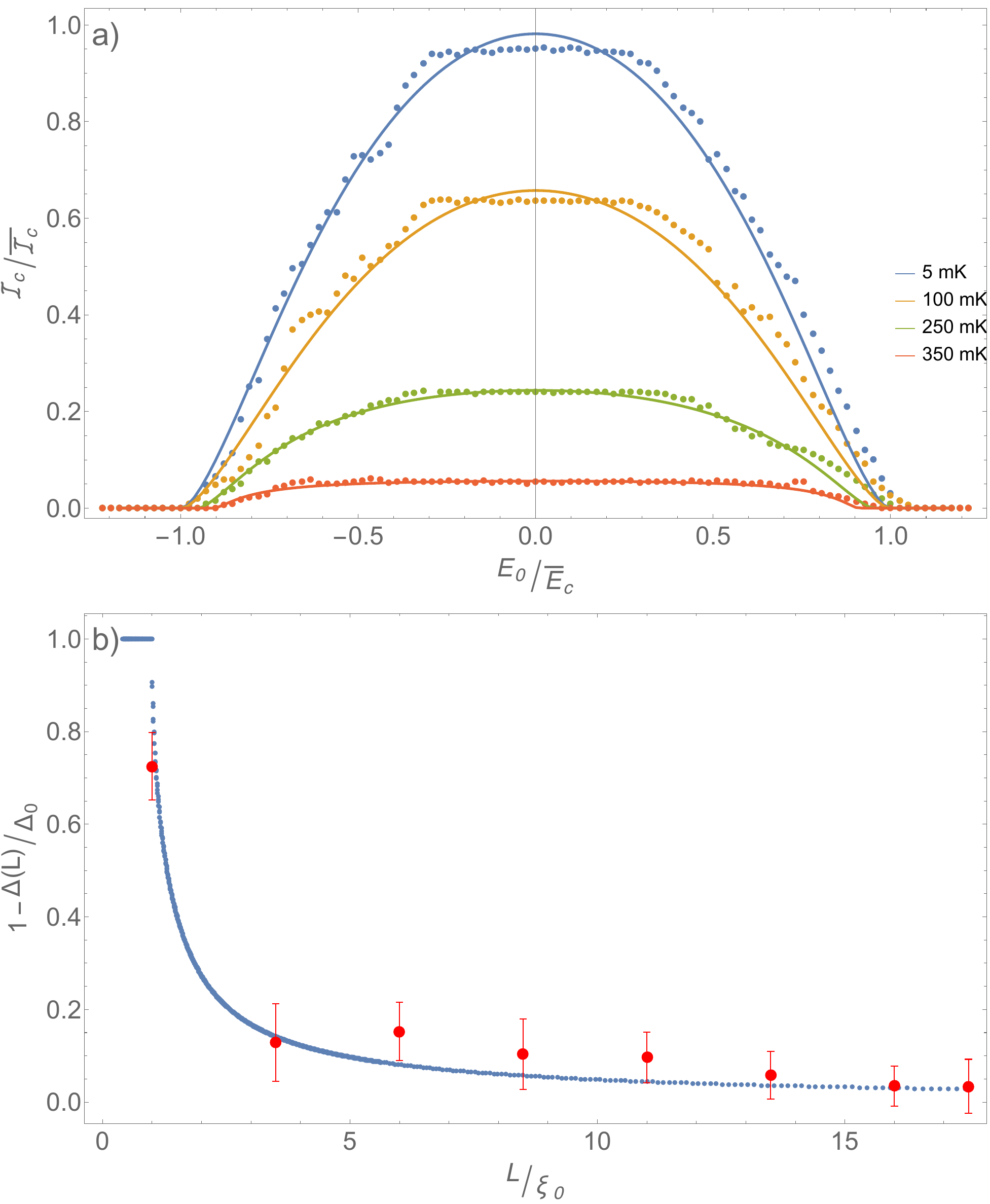}
		\caption{\textbf{a)}:  Numerical computation of the averaged critical current $\mathcal{I}_{\mathrm{c}}$ normalized against its values at $T=5$ mK, $\bar{\mathcal{I}_{\mathrm{c}}}$, as a function of the applied electric field $E_{0}$ (see \eqref{Eq:electric}) normalized against the 5 mK critical electric field $\bar{E}_{\mathrm{c}}$. 
			The curves are the numerical simulations and the dots are the experimental data from \cite{DeSimoni_2018}.
			The values of $\bar{\mathcal{I}_{\mathrm{c}}}$ at different temperatures are taken from the experiments.
			The parameters $\beta_1$ and $\beta_{2}$ are fixed as discussed in the text.
			\textbf{b)}: The suppression of the electric field effect as a function of the thickness of the film $L$ for $T=5$ mK and $E=\bar{E}_{\mathrm{c}}$.
			The red dots are the experimental data from \cite{DeSimoni_2018}.
		}
		\label{Fig:gapflowwithE}
	\end{figure}

	We are now in position to discuss the numerical results derived from the solution of the full GL equation \eqref{Eq:gapequation}, which are presented in fig.~\ref{Fig:gapflowwithE}. Fig.~\ref{Fig:gapflowwithE} a) displays the critical current of the wire, i.e.~the maximal current that the superconductor can sustain \cite{tinkham2004introduction}, $\mathcal{I}_{\mathrm{c}}$ against the applied electric field at the material boundaries, $E_{0}$, for various temperatures. In GL theory, $\mathcal{I}_{\mathrm{c}} \propto \Delta^{3}$; in our inhomogeneous situation, we take $\mathcal{I}_{\mathrm{c}} \propto (\Delta_{\mathrm{av.}})^{3}$ \footnote{$(\Delta_{\mathrm{av.}})^3$ is almost equivalent to $(\Delta^3)_{\mathrm{av.}}$, as the profiles are weakly dependent on $z$.}. As displayed, the presence of a strong electric field weakens superconductivity and eventually leads to a vanishing $\mathcal{I}_{\mathrm{c}}$ which corresponds to a superconducting to normal phase transition (i.e.~$\Delta(z) \rightarrow 0$). The numerical simulations reproduce the qualitative behaviour of the experiments \cite{DeSimoni_2020} at low temperatures. This should be expected, since the GL approach to superconductivity is expected to be accurate for $T \lesssim T_{\mathrm{c}}$. For larger temperatures, our GL model is only expected to reproduce the qualitative features of superconductive transport, as seen in Fig.~\ref{Fig:gapflowwithE} a).

	From the numerical data for $\mathcal{I}_{\mathrm{c}}$ we observe a small asymmetry in the critical electric field between positive and negative choices for the electrodes. This is evident from the equation of motion for the gap \eqref{Eq:gapequation} as the chemical potential term, $q \varphi$, is not symmetric under $E_{0} \rightarrow -E_{0}$. Such an asymmetry also seems to be present in the experimental data, but the effect is significantly stronger than what we predict. It remains to be demonstrated experimentally whether this is a real effect, as this would point to additional terms missing from our modified GL free energy \eqref{Eq:Freenergy}.
	
	An important feature of the experimental data which has not yet been explained in the literature is the emergence of the coherence length scale in the dependence of the SFE on film thickness. To examine this issue, we set $T=5$ mK and $E_{0} = \bar{E}_{\mathrm{c}} \simeq 10^8$ V m$^{-1}$ and consider films with different thicknesses $L$. Plots for other temperatures are qualitatively similar. The numerical results are sketched in Figure \ref{Fig:gapflowwithE} b). For $L \lesssim \xi_0$ the electric field causes a complete phase transition to the metallic phase. As $L > \xi_0$ is increased, one finds that the same external electric field does not completely suppress the gap. The effect becomes completely irrelevant if the sample is thicker than $7-8$ coherence lengths. This behaviour matches the experimental observations \cite{DeSimoni_2018,Alegria_2021,Ritter_2021} (shown with the red dots in the figure).
	
	This latter observation strongly supports the idea that the SFE emerges from an interplay between the small electric field penetration length $\lambda_E$ and the much larger superconducting coherence length $\xi_0$. This has an interesting interpretation in terms of the renormalization group. The naive scaling dimension of the couplings $\beta_1$ and $\beta_2$ shows that these terms are marginal in $d=2$ dimensions and irrelevant for $d>2$. When $L\simeq\xi_0$, and the material is effectively two dimensional, there can be a phase transition as the $\beta$-corrections are important for defining the low energy behaviour of the Effective Field Theory (EFT). This numerical and theoretical analysis also explains why these kinds of phase transition have never been seen in extended three dimensional samples.
	
	\begin{figure}
		\centering	
		\includegraphics[scale=.26]{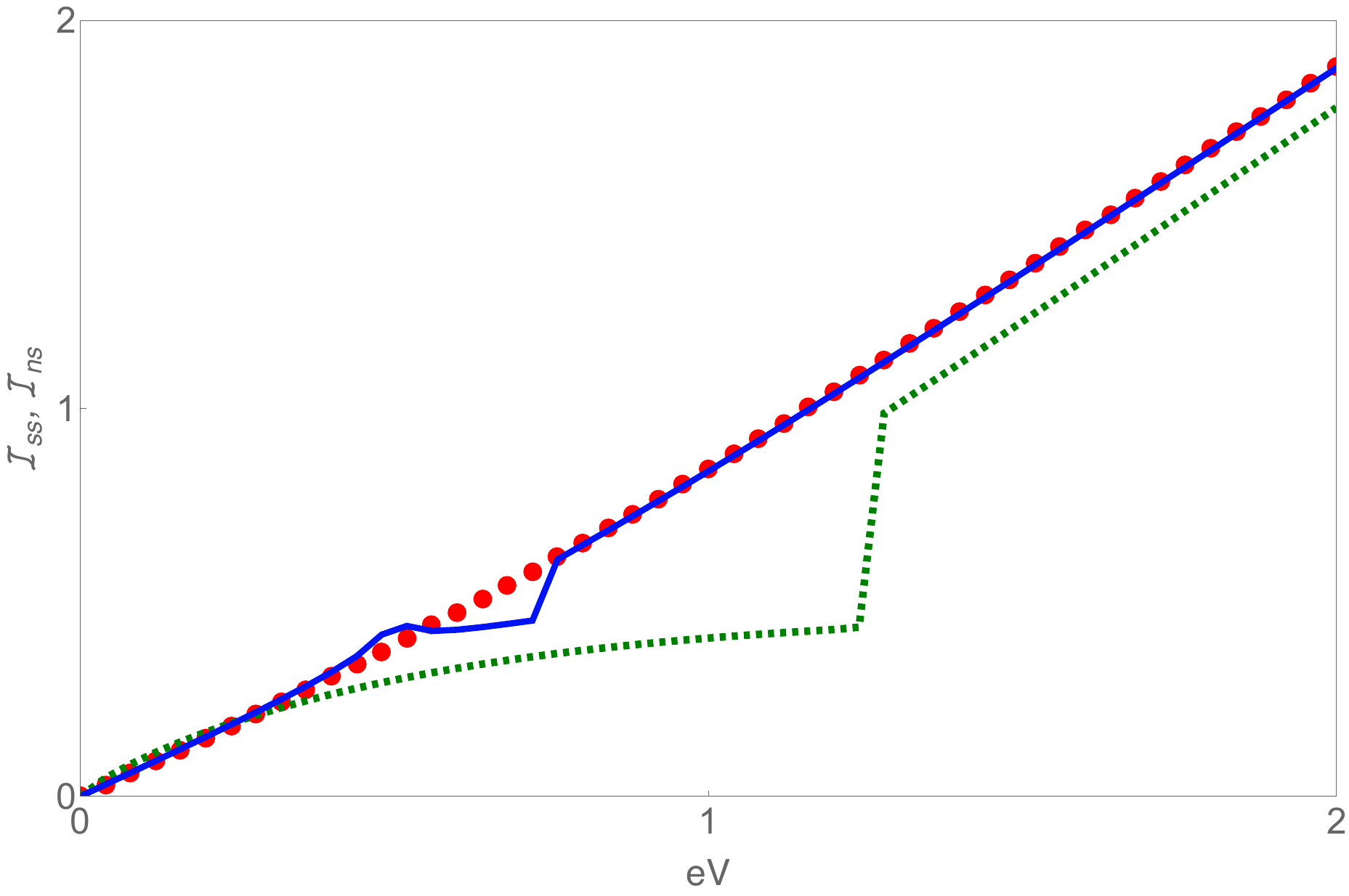}
		\caption{Tunneling currents as a function of the applied voltage $eV$ through the junction for $E_{0}/\bar{E}_{\mathrm{c}}=0$ (dashed green curve), $E_{0}/\bar{E}_{\mathrm{c}}=0.98$ (blue curve) and $E_{0}/\bar{E}_{\mathrm{c}}=1.1$ (red dots line). The temperature is $T=250$ mK $\simeq 3 T_{\mathrm{c}}/5$.}
		\label{Fig:tunneling}
	\end{figure}

 \emph{Electron tunneling \textemdash} 
Up to this point we have compared predictions of our modified GL model, \eqref{Eq:Freenergy} and \eqref{Eq:Permittivitydependence}, with the published experimental data.
Here, we discuss what the presented theory predicts in a tunneling experiment.

The tunneling between two different superconductors separated by a tunnel junction gives access to information about the density of states $N(\mathcal{E})$ \cite{tinkham2004introduction}, where $\mathcal{E}$ is the energy.
In this case one can think about two pieces of identical superconducting material separated by an insulating barrier, with the electric field applied only to one side.
We assume that the superconductor excitations are locally in a Fermi distribution. Furthermore, we expect the functional dependence of the density of states to be very similar to the standard BCS case. This follows from the expectation that any deviations from the BCS density of states are induced, in this case, by inhomogeneities in $\Delta$. Said inhomogeneities however are, in our situation, small - order 1 $\%$ - and appear only near the boundaries of the sample (see fig.~\ref{Fig:data1}a in the Appendix). As such, only
negligible deviations from the form of the $\Delta(z) = \Delta_0$ density of states are expected \footnote{This has also been observed in stronger deviations of $\Delta$ from its constant value, as around vortical defects \cite{PhysRevLett.62.3089}.} and thus we assume that $N(\mathcal{E}) = |\mathcal{E}| / \sqrt{|\mathcal{E}^2-\Delta^2|}$. The applied electric field $E$ does however modify the gap ($\Delta=\Delta_{\mathrm{av.}}(E_0)$) and thus the value, if not the form, of the density of states.

When the superconductor-superconductor junction is subject to a external voltage $V$, the tunneling current is given by \cite{tinkham2004introduction}:
\begin{eqnarray}\label{Eq:Iss}
	I_{ss}\propto\int_{-\infty}^{\infty}&&\frac{|\mathcal{E}|}{[\mathcal{E}^2-\Delta_{\mathrm{av.}}(E_0)^2]^{1/2}} \frac{|\mathcal{E}+eV|}{[(\mathcal{E}+eV)^2-\Delta_{0}^2]^{1/2}}\times \nonumber \\
	&&\left[f(\mathcal{E})-f(\mathcal{E}+eV)\right] d\mathcal{E} \ ,
\end{eqnarray}
where $f(\mathcal{E})$ is the Fermi distribution at energy $\mathcal{E}$. The numerical results from our model at $T=250$ mK are shown in Figure \ref{Fig:tunneling}.
The tunneling current is weakly modified for electric field up to $0.9~\bar{E}_{\mathrm{c}}$ \cite{tinkham2004introduction}.
However, for stronger electric fields, the superconductor gap decreases and new features emerge.
The discontinuous jump at voltages of $eV=\Delta_{\mathrm{av.}}(E_0)+\Delta_{0}$ (present for all values of $E_{0}$) shifts towards lower voltages and a new peak at $eV=|\Delta_{\mathrm{av.}}(E_0)-\Delta_{0}|$ appears.
When the electric field exceeds $\bar{E}_{\mathrm{c}}$, a phase transition occurs and one side of the junction becomes metallic.
The expected current $\mathcal{I}_{ns}$ \cite{tinkham2004introduction} is shown in fig.~\ref{Fig:tunneling}.

A similar tunneling experiment has been proposed in Ref. \cite{Mercaldo2021}.
In that case, the interaction between the superconductor and the electric field was mediated by the orbital degrees of freedom of the material. Because of the different underlying theories, the tunneling experiment predictions were quantitatively different from ours.
Indeed, a high accurate experiment could discriminate between the two theories.

In this direction, an interesting experimental setup was used in Ref. \cite{Alegria_2021}.
However, the experiment was not designed to measure the change in the matching peak and the superconductor-normal metal transition so it is likely that the effects discussed in Fig. \ref{Fig:tunneling} were not accessible.

\emph{Conclusions \textemdash} 
In this paper we have shown that, even when the screening effects are taken into account, a sufficiently strong electric field leads to the breakdown of superconductivity and the emergence of a normal-metal phase. Theoretically, this phase transition is induced by the mutual backreaction between the order parameter and the electric field, encoded in the modification of the electric permittivity appearing in the GL functional. The existence of these couplings is perfectly justified in an EFT approach, as they respect all the basic symmetries of the original GL functional.  In \cite{Prange_1963}, small corrections to the electric permittivity due to the superconductive energy gap were already found in the context of BCS theory and their structure is exactly the same as the interactions considered in this paper. Differently from our approach however,  \cite{Prange_1963} relies on a small electric field approximation, and the numerical values can not be compared. In
addition, \cite{Prange_1963} presumes a fixed ground state with a constant gap, while we consider an in general inhomogeneous ground state. The precise values of the new couplings have been obtained phenomenologically, by matching our theory against recent experimental observations in \cite{DeSimoni_2018,Alegria_2021,Ritter_2021}. 

Without any further assumption or fitting parameter,  we found that: $i)$ consistent with an EFT approach, parametrically small corrections to the electric permittivity are sufficient to drive a phase transition and to match the measured behavior of the critical current as a function of the applied electric field,
$ii)$ the qualitative behaviour is independent of the charge sign of the electrodes, $iii)$ the critical temperature is not appreciably affected by the electric field applied, $iv)$ the effect vanishes when the thickness of the sample is increased over several coherence lengths.
All these predictions are both qualitatively and quantitatively in agreement with the experiments
\cite{DeSimoni_2018,Paolucci_2018,Paolucci_2019,Paolucci_2019_simm,Paolucci_2019_squid,Paolucci2021_ionic,Alegria_2021,Ritter_2021}.
Additionally, we have proposed a way to further test our theoretical model through electron tunneling, highlighting the features that could be detected in these kind of experiments. 

The success of the present model in predicting the observed experimental behaviour opens up several new directions for further research.
Arguably the most important task is to calculate the additional couplings $\beta_1$ and $\beta_2$ directly from a microscopic theory. Additionally, it would be interesting to analyze whether modulating the phase of the condensate alters our results. This can be done by, e.g., considering a SQUID setup and attempting to match the results to the data of \cite{Paolucci_2019_squid,DeSimoni2021_SQUID}.

\begin{acknowledgments}
	\section{Acknowledgments}
	The authors are listed alphabetically. We thank Francesco Giazotto and Giorgio De Simoni for sharing the data of \cite{DeSimoni_2018} and for insightful discussions.   AA and IM have been partially supported by the ``Curiosity Driven Grant 2020'' of the University of Genoa. The authors have been partially supported by the INFN Scientific Initiatives SFT: ``Statistical Field Theory, Low-Dimensional Systems, Integrable Models and Applications'' and BELL. This project has also received funding from the European Union's Horizon $2020$ research and innovation programme under the Marie Sk\l{}odowska-Curie grant agreement No. $101030915$.
\end{acknowledgments}
	
\widetext
\appendix
\setcounter{equation}{0}
\setcounter{table}{0}
\makeatletter
\renewcommand{\theequation}{A.\arabic{equation}}
	\section{Appendix: technical and numerical aspects of the computation of the order parameter}
	\label{app}
	We employ the following normalisations to render the equations of motion dimensionless (and thus suitable for numerical calculations). Let $L$ be the width of the sample and $(x,y,z)$ the spatial coordinates so that:
	\begin{subequations}
		\begin{eqnarray}
			(x,y,z) \rightarrow (u,v,w) &=& \frac{1}{L} (x,y,z) \; , \\
			\Delta(z) \rightarrow \bar{\Delta}(w) &=& \sqrt{-\frac{2 \alpha_{4}}{3 \alpha_{2}}} \Delta(w) \; , \\
			E_{z} \rightarrow \bar{E}_{z} &=& \frac{q L^2}{c \hbar} E_{z} \; , \\
			\alpha_{2} \rightarrow \bar{\alpha}_{2} &=& -\frac{m L^2}{\hbar^2} \alpha_{2} \; , \\
			\alpha_{4} \rightarrow \bar{\alpha}_{4} &=& \frac{m L^2 \Delta_{0}^2}{\hbar^2} \alpha_{4} = -\frac{3 m L^2}{2 \hbar^2}  \alpha_{2} \; , \\
			\epsilon_{0} \rightarrow \bar{\epsilon}_{0} &=& \frac{m c^2  \Delta_{0}^2}{q^2 L^2} \epsilon_{0} = -\frac{2 m c^2 \alpha_{4}}{3 \alpha_{2} q^2 L^2} \epsilon_{0} \; , \\
			\beta_{1} \rightarrow \bar{\beta}_{1} &=& \frac{m \epsilon_{0} c^2}{2 q^2 L^2} \beta_{1} \; , \\
			\beta_{2} \rightarrow \bar{\beta}_{2} &=& \frac{m \epsilon_{0} c^2 \Delta_{0}^2}{2 q^2 L^2} \beta_{2} = -\frac{3 m \epsilon_{0} c^2 \alpha_{2}}{4 q^2 L^2 \alpha_{4}} \beta_{2} \; .
		\end{eqnarray}
	\end{subequations}
	As all our solutions will only depend on $z$ we can drop all the derivatives with respect to $u$ and $v$. Subsequently, after these replacements, the equation of motion for the phase becomes
	\begin{eqnarray}
		\label{Eq:phaseom}
		0 &=& \frac{d}{d w} \left( \bar{\Delta}^2 \frac{d}{d w} \theta \right) \; ,
	\end{eqnarray}
	As is standard we will choose a constant phase profile which solves \eqref{Eq:phaseom} identically. This choice also minimises the free energy contribution of this corresponding term. Hence the equation of motion for the gap becomes
	\begin{eqnarray}
		\label{Eq:eom}
		0 &=&\frac{d^{2}}{d w ^2} \bar{\Delta} + \left( \bar{\alpha}_{2} - \left( \frac{2 m c L^2}{\hbar} \bar{\varphi} + \bar{\beta}_{1} \left( \frac{d}{d w}  \bar{\varphi} \right)^2  \right) \right) \bar{\Delta} \nonumber \\
		&\;& - 
		\left( \frac{2 \bar{\alpha}_{4}}{3} + 2 \bar{\beta}_{2} \left( \frac{d}{d w}  \bar{\varphi} \right)^2 \right) \bar{\Delta}^3 \; .
	\end{eqnarray}

	\begin{figure}[t]
		\centering
		\includegraphics[width=0.45\linewidth]{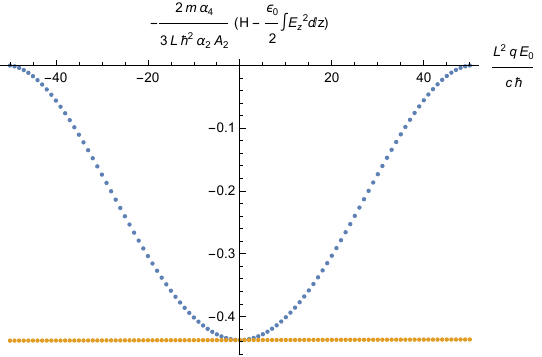}
		\caption{A plot of the superconductor energy at zero temperature (minus a constant permittivity contribution) against the applied electric field. The blue dots represent our model while the yellow dots are the standard Ginzburg-Landau model ($\epsilon [\Delta] = \epsilon_{0}$) with a chemical potential. The normal phase has zero energy (modulo the constant permittivity term) and lies along the horizontal axis. We can see as the electric field is increased the energy of our system eventually becomes equal to that of the energy of the normal phase indicating a phase transition.}
		\label{Fig:energy}
	\end{figure}
	
	We use the standard Mathematica routine NDSolve to integrate \eqref{Eq:eom} from the boundary of the material at $w=0$ to $w=1$ where again $w=z/L$ with $L$ the width of the superconductor.  The boundary condition on the condensate requires that we set $\frac{d \bar{\Delta}}{d w}(0)  =0$ but leaves the initial value, $\bar{\Delta}(0)$, unfixed. As such we use this initial value as a free parameter and integrate from $w=0$ to $w=1$ with some choice of this parameter. We search for solutions where  $\frac{d \bar{\Delta}}{dw}\left(\frac{1}{2}\right)=0$ by adjusting the value of $\bar{\Delta}(0)$ (i.e.~we use a shooting method). This second condition ensures that the condensate is symmetric about the midpoint of the material. Subsequently we confirm that $\frac{d \bar{\Delta}}{dw}(1)=0$ as a crosscheck that the solution is symmetric about the midpoint.
	
	\begin{figure}[t]
		\centering
		\subfloat[][]{\includegraphics[width=0.45\linewidth]{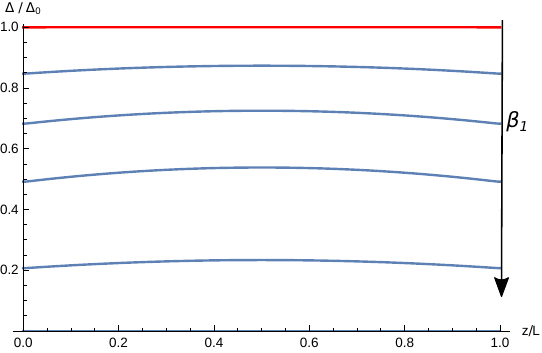}
			\label{fig:constantpermittivity}} \qquad
		\subfloat[][]{\includegraphics[width=0.45\linewidth]{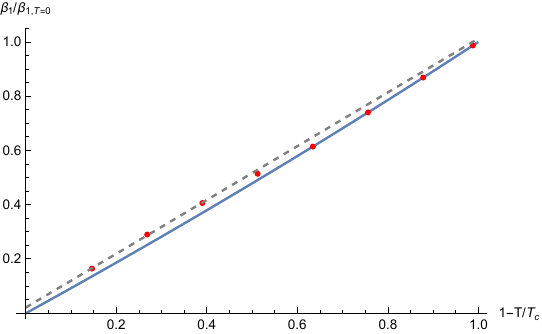}
			\label{fig:dependenceofbeta1onT}}	
		\caption{\protect\subref{fig:constantpermittivity} The profile of the gap in the $z$-direction at the critical electric field $\bar{E}_{\mathrm{c}}$ and zero temperature with various values of the coupling $\beta_{1}$. The red line corresponds to the constant permittivity case ($\epsilon[\Delta]=\epsilon_{0}$) and depends only weakly on the position. For each blue line of smaller $\bar{\Delta}(z)$ we are increasing $\beta_{1}$ (from $1.42 \times 10^{-29}$ m$^3$ to $1.13 \times 10^{-28}$ m$^3$ in steps of $1.42 \times 10^{-29}$ m$^3$). \protect\subref{fig:dependenceofbeta1onT}  The dependence of $\beta_{1}$ on $T/T_{\mathrm{c}}$ normalized by the value of $\beta_{1}$ at zero temperature (given in Eq. \eqref{Eq:beta1val}). The blue line represents how $\beta_{1}$ must change if the critical electric field is unaffected by temperature. The red points are the critical electric field extracted from experimental data \cite{DeSimoni_2018}. The grey dashed line is a linear fit to the first three experimental points and it has a non-zero, positive intercept with the $y$-axis.}
		\label{Fig:data1}
	\end{figure}
	
	To confirm that we are indeed finding a minimal energy solution at zero temperature we can compute the superconductor energy. In terms of our dimensionless quantities the energy has the form
	\begin{eqnarray}
		- \frac{2 m \alpha_{4}}{3 L \hbar^2 \alpha_{2}} H &=& A_{2} \int dw \; \Big \{\frac{1}{2} \left( \frac{d \bar{\Delta}}{dw} \right)^2 + \left( - \frac{\bar{\alpha}_{2}}{2} +  \frac{m c L^2}{\hbar} \bar{\varphi} \right) \bar{\Delta}^2 +  \frac{\bar{\alpha}_{4}}{4!}  \bar{\Delta}^4 \nonumber \\
		&\;& \hphantom{\int du dv dw \; \Big \{} + \frac{1}{2} \left( \bar{\epsilon}_{0} + \bar{\beta}_{1} \bar{\Delta}^2 + \bar{\beta}_{2} \bar{\Delta}^4 \right) \left( \frac{d \bar{\varphi}}{dw} \right)^2  \Big \} \; ,
	\end{eqnarray}
	with $A_{2}$ the area of the sample in the $u$ and $v$-directions. We plot its value (minus the contribution from the constant permittivity term) against the applied electric field in fig.~\ref{Fig:energy}.  For the displayed quantity the normal phase, which corresponds to $\bar{\Delta}(w) = 0$, lies along the horizontal axis. Consequently we see that our solution has a lower energy than the normal state for small to moderate electric fields. As we approach the critical electric field the energy of the ground state increases until it is comparable with the normal state signalling a phase transition.
	
	\begin{figure}[t]
		\centering
		\subfloat[][]{\includegraphics[width=0.45\linewidth]{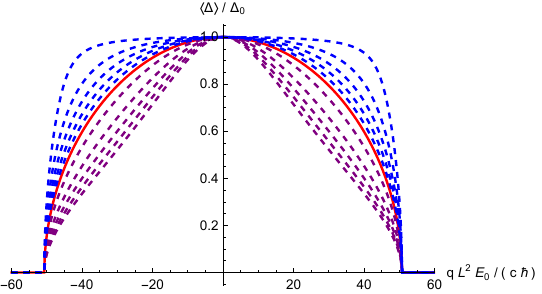}
			\label{fig:condensateflowagainstEvariousbeta2}} \qquad	
		\subfloat[][]{\includegraphics[width=0.45\linewidth]{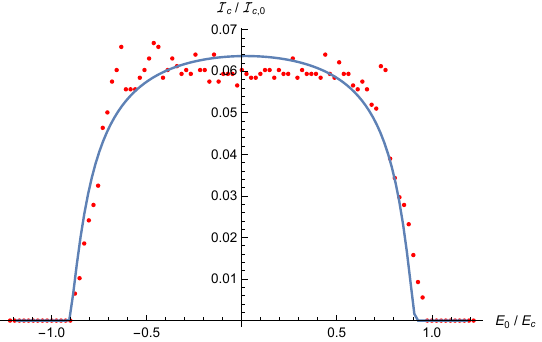}
			\label{fig:350mK}}
		\caption{\protect\subref{fig:condensateflowagainstEvariousbeta2}  The dependence of the average gap at zero temperature on the electric field applied at the edge of the superconductor for $\beta_{1}$ having the critical value \eqref{Eq:beta1val}. The red line corresponds to $\beta_{2}=0$. The purple dashed lines indicate values of $\beta_{2}$ greater than zero (below the red line) from $1.97 \times 10^{-54}$ m$^6$ to $9.83 \times 10^{-54}$ m$^6$ in steps of $1.97 \times 10^{-54}$ m$^6$. The blue dashed lines are for values of $\beta_{2}$ less than zero (above the red line) from $-3.93 \times 10^{-55}$ m$^6$ to $-1.97 \times 10^{-54}$ m$^6$ in steps of $-3.93 \times 10^{-55}$ m$^6$. \protect\subref{fig:350mK} A plot of the critical current, normalised by its zero temperature value, against the applied electric field. The red dots are data while the solid blue line is a best fit from our model.}
		\label{Fig:data2}
	\end{figure}
	\subsection{Constant permittivity}
	We first consider the situation at zero temperature with a constant permittivity ($\epsilon[\Delta]=\epsilon_{0}$). In this case, it is still possible to drive a phase transition with the electric field; however, this requires an applied electric field of
	\begin{eqnarray}
		E_{0} \approx 5.162 \times 10^{11} \mathrm{\; V m^{-1}}
	\end{eqnarray}
	- several orders of magnitude larger than what is used in the experiments of \cite{DeSimoni_2018} where $\bar{E}_{\mathrm{c}} \approx 10^8$ Vm$^{-1}$. Moreover, for $E_{0}$ of the opposite sign there is no phase transition at all. This is because the chemical potential term in \eqref{Eq:eom} is not symmetric under $E_{0} \leftrightarrow - E_{0}$ and only one choice of sign leads to a reduction in the gap. The order of magnitude of the electric field and the unipolarity of this effect are in contrast with what has been observed in the experiments. Therefore, we need to include the change of the permittivity as a function of the superconducting gap.
	
	The profile of the gap at the critical electric field $E_{z} = \bar{E}_{\mathrm{c}} = 10^8$ Vm$^{-1}$ and constant permittivity is given by the red line in Fig.~\ref{fig:constantpermittivity}. Despite appearances, the electric field does lead to a position dependent condensate in this case but the dependence is too weak to be visible.
	
	\subsection{Quadratic density correction to the permittivity }
	Now we consider quadratic density corrections to the permittivity so that $\epsilon[\Delta] = \epsilon_{0} ( 1 + \beta_{1} \Delta^2 )$. Firstly we work at zero temperature. The red line in fig.~\ref{fig:condensateflowagainstEvariousbeta2} demonstrates the effect of a non-zero $\beta_{1}$ on the average gap against the applied electric field (at zero temperature). The value of $\beta_{1}$ in this image is chosen so that there is a phase transition at the experimental value of the critical electric field $\bar{E}_{\mathrm{c}} = 10^8$ Vm$^{-1}$ , i.e.,
	\begin{eqnarray}
		\label{Eq:beta1val}
		\beta_{1} &\approx& 5.996 \times 10^{-29} \mathrm{\; m^3} \; . 
	\end{eqnarray}
	Plots of the gap against position for various other values of $\beta_{1}$ less than this critical value, but at fixed electric field $\bar{E}_{\mathrm{c}}$, are given by the blue lines in Fig.~\ref{fig:constantpermittivity}. As $\beta_{1}$ is increased towards the critical value we can see that the gap near the edges of the superconductor decreases as argued in the main text. The profiles also keep a small gradient as expected to minimize the kinetic energy; as such the gap in the interior is forced to decrease from unity even though the electric field is effectively restricted to the skin of the material.
	
	At non-zero temperature we need to account for the fact that the critical electric field appears to be weakly dependent on the temperature. With the dependence of $\alpha_{2}$ and $\alpha_{4}$ on temperature this leads to $\beta_{1}$ having a non-zero dependence on temperature. This is displayed in Fig.~\ref{fig:dependenceofbeta1onT}. In what follows, whenever it is necessary to determine the value of $\beta_{1}$ at a given temperature, we shall use an interpolation between the experimental points with the addition of one point at zero temperature (where the $\beta_{1}$ coupling takes the previously obtained value \eqref{Eq:beta1val}).
	
	With the temperature dependence of $\beta_{1}$ fixed one can determine the dependence of the average gap on the applied electric field. Before doing this however we shall consider quartic corrections in the density to the permittivity as these help to give more realistic profiles.
	
	\subsection{Quartic density correction to the permittivity }
	In general one can consider very general density dependent corrections to the permittivity. As an example of the effect that higher density corrections can have on the average gap, in the main text we have included $\beta_{2}$ so that $\epsilon[\Delta] = \epsilon_{0} ( 1 + \beta_{1} \Delta^2 + \beta_{2} \Delta^4 )$. In Fig.~\ref{fig:condensateflowagainstEvariousbeta2} the dashed blue and purple lines represent the effect of $\beta_{2}$ at zero temperature (with $\beta_{1}$ fixed to the critical zero temperature value \eqref{Eq:beta1val}) and the red line gives the dependence of the average gap for $\beta_{2}=0$. We can see that as $\beta_{2}$ is increased the average gap is decreased; meanwhile if we take $\beta_{2}$ negative the average gap is increased and the curve becomes less sharp about $E_{0}=0$. As such we can tune $\beta_{2}$ negative to generate a more flat profile about $E_{0}=0$ to better match experimental results. 
	
	\begin{figure*}[t]
		\centering
		\includegraphics[width=0.45\linewidth]{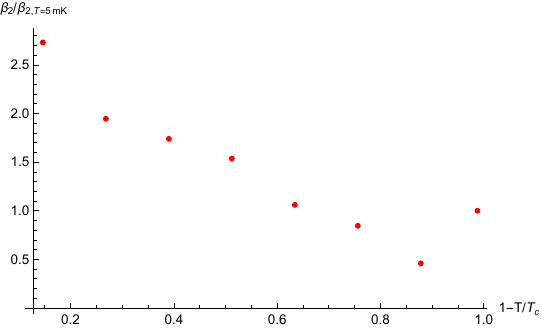}
		\caption{The dependence of the $\beta_{2}$ coupling, normalised by the value at 5 mK ($-7.27 \times 10^{-55}$ m$^6$) against temperature extracted by fitting our model to experimental data.}
		\label{fig:beta2flowwithT}
	\end{figure*}
	
	At non-zero temperature we compute the mean-square-difference between our model profiles and the experimental data. We then minimise this measure to obtain the optimal value of $\beta_{2}$ at each temperature. The resultant dependence of the average gap on the electric field for $T=350$ mK is displayed in Fig.~\ref{fig:350mK}. The values of $\beta_{2}$ for other temperatures are given in Fig.~\ref{fig:beta2flowwithT}.
	
	\subsection{Kinetic, potential and boundary energies}
\begin{figure*}[t]
		\centering
		\subfloat[][]{\includegraphics[width=0.45\linewidth]{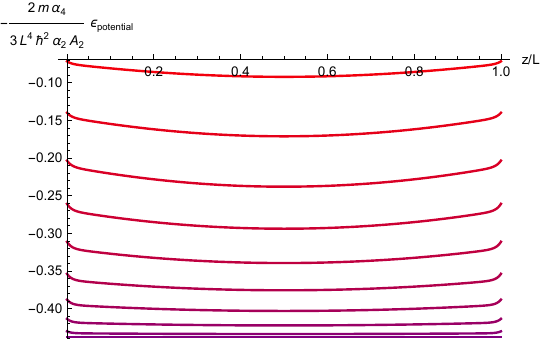}
			\label{fig:potentialpos}} \qquad
		\subfloat[][]{\includegraphics[width=0.45\linewidth]{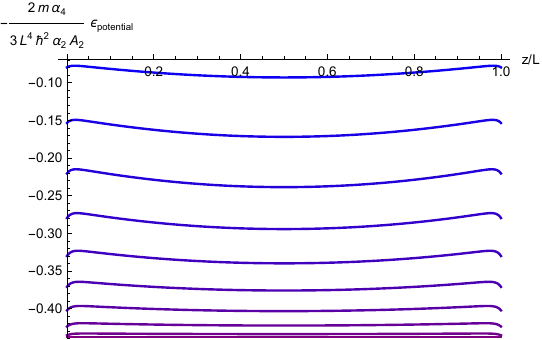}
			\label{fig:potentialneg}} \\
		\subfloat[][]{\includegraphics[width=0.45\linewidth]{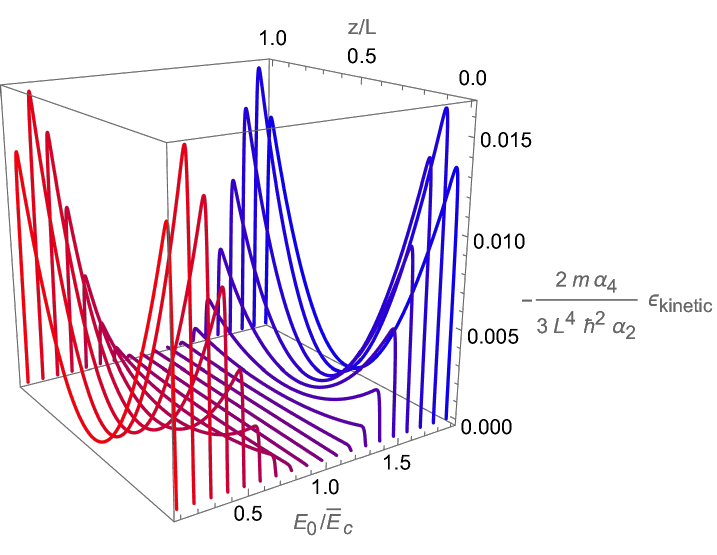}
			\label{fig:kinetic}} \qquad
		\subfloat[][]{\includegraphics[width=0.45\linewidth]{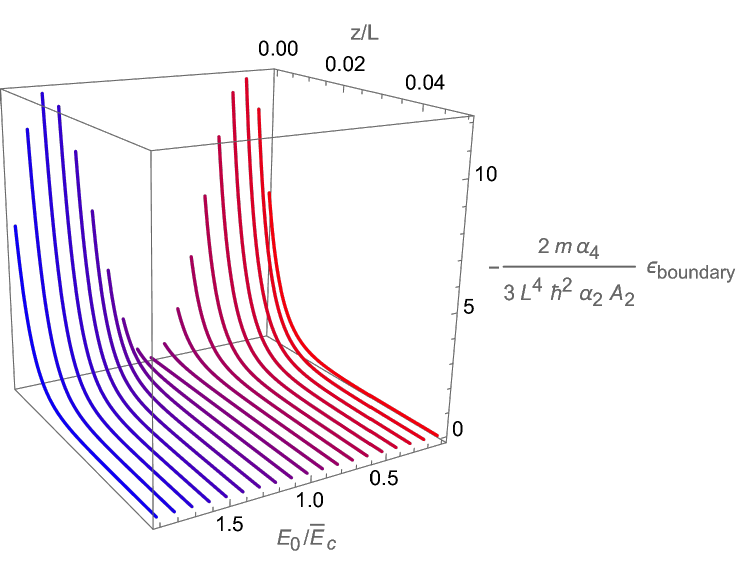}
			\label{fig:boundary}} 
		\caption{Plots of the energy densities. Each line is a different choice of $E_{0}/E_{\mathrm{c},T=0}$ with red corresponding to positive values and blue to negative values of $E_{0}$. \protect\subref{fig:potentialpos} is a plot of the potential energy density against position for $E_{0}>0$ with $E_{0}/E_{\mathrm{c},T=0} = 0, 1/10, \ldots, 9/10$ (bottom to top). \protect\subref{fig:potentialneg} is a plot of the potential energy density against position for $E_{0}<0$ with $E_{0}/E_{\mathrm{c},T=0} = 0, -1/10, \ldots, -9/10$ (bottom to top). \protect\subref{fig:kinetic} and \protect\subref{fig:boundary} represent the values of the kinetic and boundary energy densities against electric field and position with the electric field ranging from $E_{0}/E_{\mathrm{c},T=0} = - 9/10, \ldots, 9/10$.}
		\label{Fig:data3}
	\end{figure*}

	The energy density of the condensate can be decomposed into three terms,
	\begin{subequations}
		\begin{eqnarray}
			\varepsilon_{\mathrm{kinetic}} &=&  \left( \frac{d}{d z}  \Delta \right)^2 \; , \\
			\varepsilon_{\mathrm{potential}} &=& \left( \frac{\alpha_{2}}{2!} + q \varphi \right) \Delta^2 +  \frac{\alpha_{4}}{4!}  \Delta^4  \; , \\
			\varepsilon_{\mathrm{boundary}} &=& \frac{\epsilon[\Delta]}{2} \left(  \frac{d}{d z}  \varphi \right)^2 \; ,
		\end{eqnarray}
	\end{subequations}
	which are the kinetic, potential and boundary energy densities respectively. In the main text we argue that the kinetic energy contribution is small compared to the other contributions for a large range of electric field values. This can be seen by looking at the values of the densities as displayed in Figs.~\ref{fig:potentialpos}, \ref{fig:potentialneg}, \ref{fig:kinetic} and \ref{fig:boundary}. The absolute value of the kinetic term is much smaller than the sum of potential and boundary energy terms.
	
	\bibliography{bib1}

\end{document}